\documentclass[twocolumn,showpacs,preprintnumbers]{revtex4}%
\usepackage{amsmath}%
\usepackage{amsfonts}%
\usepackage{amssymb}%
\usepackage{graphicx}
\begin{document}
\title{Variation of the glass transition temperature with rigidity and chemical composition}
\author{Gerardo G. Naumis}
\affiliation{Instituto de F\'{\i}sica, Universidad Nacional Aut\'{o}noma de M\'{e}xico
(UNAM), Apartado Postal 20-364, 01000, M\'{e}xico, Distrito Federal, Mexico.}
\date{\today }


\begin{abstract}
The effects of rigidity and chemical composition in the variation of the glass
transition temperature are obtained by using the Lindemann criteria, that
relates melting temperature with atomic vibrations. Using this criteria and
that floppy modes at low frequencies enhance in a considerable way the average
cuadratic displacement, we show that the consequence is a modified glass
transition temperature. This approach allows to obtain in a simple way the
empirically modified Gibbs-DiMarzio law, which has been widely used in
chalcogenide glasses to fit the changes in the glass transition temperature
with the chemical composition . The method predicts that the constant that
appears in the law depends upon the ratio of two characteristic frequencies
(or temperatures). Then, the constant for the $Se_{1-x-y}Ge_{x}As_{y}$ glass
is estimated by using the experimental density of vibrational states, and the
result shows a very good agreement with the experimental fit from glass
transition temperature variation.

\end{abstract}
\maketitle

\address{$^{1}$Instituto de Fisica, Universidad Nacional Aut\'{o}noma de
M\'{e}xico (UNAM)\\
Apartado Postal 20-364, 01000, Distrito Federal, Mexico.}

Glasses are solids that do not have long range order, and which are usually
produced by fast cooling of a liquid melt. In spite of the well known
importance of these materials in human history, the physics of glass formation
is still an open and puzzling problem \cite{Anderson}. The reason behind this
situation is that glass formation is mainly a non-equilibrium process, and
thus even some important questions for glass technology are not well
understood, as the origin of the non-exponential relaxation laws
\cite{Relaxation} or the ability of certain materials to reach the glassy
state \cite{Jackle}. The first approach to the problem was made by Kauzmann
\cite{Debenebook}, \ who pointed out that there is an underlying
thermodynamical phase transition behind glass formation due to an entropy
crisis. Although the Kauzmann approach was very successful, it was unable to
answer many questions, specially in a quantitative way. Later on, other
approaches were proposed \cite{Debenedetti}: phenomenological models like the
Gibbs-Dimarzio, theories based in supercooled liquids as for example mode
coupling and energy landscape formalism, or the use of extensive computer
simulations \cite{Debenebook}. \ However, most of these theories have
difficulties to explain one of the most simple and interesting questions: how
the glass transition temperature ($T_{g}$) depends on chemical composition. As
discovered by the Phoenicians more than 2,000 years ago, $T_{g}$ can be
dramatically lowered or raised by adding few impurities. Another interesting
property of glasses is the behavior of viscosity, which is usually referred as
fragility \cite{Tatsumisago}. Fragility is related to the glass forming
tendency, since a non-fragile glass former do not require a high speed of
cooling to be produced. The fragility of the glass can be changed from strong
to fragile \cite{Tatsumisago}. Chalcogenide glasses (formed with elements from
the VI column doped with impurities) are very important to understand these
effects. In general, the change of $T_{g}$ with the chemical composition has
been observed to follow an empirically modified Gibbs-DiMarzio law
\cite{Tatsumisago}\cite{Sreeram}\cite{Boolchand1}\cite{GeorgievCR}. It is
worthwhile mentioning that the original Gibbs-DiMarzio law was developed for
polymers \cite{Debenebook}, and contains some special assumptions that have
been very elusive to measure, like the size of cooperative rearranging regions
\cite{Debenebook}. Furthermore, the use of this law for chalcogenide glasses
is based in the hypotesis that these glasses nearly behave as polymers. A
completely different approach is to use stochastic matrices
\cite{Kerner1,Kerner2}\cite{Micoulaut}. The problem with this theory is that
one needs to assume certain conditions to define the glass transition which
are not in the form of usual definitions \cite{Kerner1}.

The rigidity theory (RT), introduced by Phillips \cite{Phillips1} and refined
by Thorpe \cite{Thorpe0}, was another important step to understand glass
formation. By considering the covalent bonding as a mechanical constraint, the
ease of glass formation is related with the ratio between available degrees of
freedom and the number of constraints. If the number of constraints is lower
than the degrees of freedom, there are zero frequency vibrational modes called
floppy \cite{Thorpe1}. The resulting network is under-constrained. A
transition occurs when a disordered lattice becomes rigid. Glasses that are
rigid at a certain chemical composition are easier to form, and many features
of this transition have been experimentally observed \cite{Boolchand1}%
\cite{Boolchand2}. Even for simple systems like hard-disks \cite{Huerta0},
polymer melts \cite{Joos}, and colloids \cite{Huerta1}, rigidity plays an
important role . For more complex systems like proteins, RT is a very powerful
tool to understand folding and long-time scale motions \cite{Brandon}.
Recently, it has been discovered that at the rigidity transition, there is a
window of thermodynamical reversibility\textit{\ }\cite{Boolchand1}%
\cite{Georgiev}\cite{Boolchand}\cite{Boolchand3}, explained as a
self-organization of the stress in the network \cite{Chubinsky}\cite{Barre}.
Although RT explains a lot of the phenomenology of chalcogenide glasses, the
introduction of thermodynamics into RT has been elusive \cite{Joos}. In
previous papers, we approached some of these problems by using a
phenomenological free energy \cite{Naumis}, and then we made computer
simulations \cite{Huerta}\cite{Huertaprb}. Yet, many questions remain open, as
for example the variation of $T_{g}$ with the chemical composition of the
glass. \ In this letter, we will show that by using the RT and the old
Lindemann criteria that relates melting with the size of atomic vibrations
(which has been proved to be valid for glasses \cite{Buchenau0}%
\cite{Buchenau1}), one can explain the variation of $T_{g}$ with chemical
composition, and obtain in a simple and elegant way the modified empirical
Gibbs-DiMarzio law. Furthermore, it provides a clear path to understand glass
transition trends with the connectivity of the network.

In RT, the ability for making a glass is optimized when the number of freedom
degrees, $3N,$ where $N$ is the number of particles, is equal to the number of
mechanical constraints ($N_{c}$) that are given by bond bending and stretching
forces. The parameter $f=(3N-N_{c})/3N$ gives the fraction of cyclic variables
of the Hamiltonian. A variation of these coordinates do not change the energy
and thus $f$ also corresponds to the fraction of vibrational modes with zero
frequency, called floppy modes \cite{Thorpe0},\ with respect to the total
number of vibrational modes. The parameter $f$ is in fact a function of the
average chemical coordination of the glass, and can be calculated using the
pebble game algorithm \cite{Thorpe1}, but usually it can be estimated with a
mean-field procedure, known as Maxwell counting \cite{Thorpe2},
\[
f=\frac{3N-N_{c}}{3N}=2-\frac{5}{6}\left\langle r\right\rangle
\]
where $\left\langle r\right\rangle $ is the average coordination number,
defined as,
\[
\left\langle r\right\rangle =\sum_{r=0}^{Z_{\max}}rx_{r}%
\]
and $r$ is the coordination number of each atomic specie. $x_{r}$ is the
occurrence of each type of atom \cite{Thorpe2}. A rigidity transition occurs
when $f=0,$ and the system passes from a floppy network to a rigid one.
Glasses are rigid if there are more constraints than degrees of freedom, and
then the lattice is overconstrained. In $3D$, the rigidity transition leads to
the critical value $\left\langle r_{c}\right\rangle =2.4$ if all angular
constraints are considered \cite{Thorpe2}. In principle, since floppy modes
have zero frequency; they do not contribute to the elastic energy, and the
specific heat of the glass depends on $f$, a result that violates the
Dulong-Petit law \cite{Naumis}, and which is not observed in the experimental
data \cite{Boolchand1} . In fact, floppy modes do not have a perfect zero
frequency, \textit{i.e.}, in real glasses they are shifted by residual forces,
like the Van der Waals interaction \cite{Thorpe0}. This is confirmed by
neutron scattering experiments, where it has been shown that floppy modes in
the prototypical compound $Se_{1-x-y}Ge_{x}As_{y}$ are blue-shifted
\cite{Boolchand2}\cite{Capeletti}\cite{Kamitakahara}, forming a peak around
$5meV.$ The corresponding temperature ($\Theta_{f}$) where these modes are
frozen, can be estimated from the energy required to excite modes of $5meV$,
that gives $\Theta_{f}\approx54%
%TCIMACRO{\U{b0}}%
%BeginExpansion
{{}^\circ}%
%EndExpansion
K.$ For low temperatures, is clear that floppy modes are important, as
confirmed by the giant-softening of the $119Sn$ Lamb-Mossbauer factor in
$(Ge_{0.99}Sn_{0.01})_{x}Se_{1-x}$ glasses \cite{Boolchand2}. But in
principle, one can argue that floppy modes are not so important for the
physics of the glass at high temperatures, since all $3Nf$ floppy oscillators
are already excited, leading to the Dulong-Petit law. As we will show next,
this is not the case, and in fact floppy modes are essential in determining
$T_{g}.$To show this, we use the Lindemann criteria (1910), which has been
originally devised to understand crystal melting \cite{Tabor}. Such criteria
establishes that melting occurs when the mean atomic square displacement
$\left\langle u^{2}(T)\right\rangle $ is around $10\%$ of the atomic spacing
$a$. There are many experimental evidences that this criteria is applicable to
glasses \cite{Buchenau1}, as for example happens with $Se$ and $Se-Ge$ systems
\cite{Buchenau0}.\ The value of $\left\langle u^{2}(T)\right\rangle $ can be
calculated from the density of vibrational states $g(\omega)$. At high
temperatures,
\begin{equation}
\left\langle u^{2}(T)\right\rangle =\frac{kT}{mN}%
%TCIMACRO{\dint \limits_{0}^{\infty}}%
%BeginExpansion
{\displaystyle\int\limits_{0}^{\infty}}
%EndExpansion
\frac{g(\omega)}{\omega^{2}}d\omega\label{usquare}%
\end{equation}
where $k$ is the Boltzmann constant, $T$ the temperature, $\omega$ the
frequency and $m$ the mass. At $T_{g},$ the Lindemann criteria applied to
glasses \cite{Buchenau0} establishes that $\left\langle u^{2}(T_{g}%
)\right\rangle \approx\left\langle u^{2}(T_{m})\right\rangle \approx0.1a^{2}$
where $T_{m}$ is the melting temperature. This shows the fundamental
importance of floppy modes to determine $T_{g}$ due to the enhancement of low
frequencies produced by the term $\omega^{-2}$ in eq.(\ref{usquare}), which
leads to an increasing $\left\langle u^{2}\right\rangle .$Let us made a model
for the most prototypical chalcogenide glass: $Se_{1-x-y}Ge_{x}As_{y}$, since
this compound allows to obtain the same $\left\langle r\right\rangle $ with
different chemical compositions. The three chemical elements of this glass
have nearly the same mass; thus we can suppose that in eq.(\ref{usquare}) all
the atomic masses are equal, although in more general cases $m$ must be
replaced by an averaged mass. The most simple form for $g(\omega)$ is to use
an Einstein like model, with a delta function centered around a characteristic
floppy peak at $\omega_{f}$, with a weight given by the number of floppy
modes$,$ plus a density of states that has the rest of the spectral weight.
Such a density of states can be written as,%
\begin{equation}
g(\omega)=3N(1-f)g_{I}(\omega)+3Nf\delta(\omega-\omega_{f})\label{gomega}%
\end{equation}
where $g_{I}(\omega)$ is a density normalized to one for $f=0$. It is
important to remark that in this model, $g_{I}(\omega)$ will be the same at
all glass compositions below $\left\langle r\right\rangle =2.4$, and the
relative weight on non-floppy modes is taken into account by the factor $1-f.
$ Note that by using a fixed $g_{I}(\omega)$ for all chemical compositions, we
are overestimating this contribution for flexible glasses, which will give
only a small correction to the final result, since $g(\omega)/\omega^{2}$ goes
to zero at high frequencies \cite{Kamitakahara}. This assumption is also
supported by the neutron scattering data \cite{Kamitakahara}, which shows that
$g(\omega)$ follows an isocoordination rule for $Se_{1-x-y}Ge_{x}As_{y}.$Using
$g(\omega)$, $\left\langle u^{2}(T)\right\rangle $ turns out to be,%
\begin{equation}
\left\langle u^{2}(T)\right\rangle =\frac{3kT}{m}\left[  \left\langle \frac
{1}{\omega^{2}}\right\rangle _{R}+f\left(  \frac{1}{\omega_{f}^{2}%
}-\left\langle \frac{1}{\omega^{2}}\right\rangle _{R}\right)  \right]
,\label{u2floppy}%
\end{equation}
where $\left\langle \frac{1}{\omega^{2}}\right\rangle _{R}$ is defined as the
second inverse moment at the rigidity transition,%
\[
\left\langle \frac{1}{\omega^{2}}\right\rangle _{R}\equiv%
%TCIMACRO{\dint \limits_{0}^{\infty}}%
%BeginExpansion
{\displaystyle\int\limits_{0}^{\infty}}
%EndExpansion
\frac{g_{I}(\omega)}{\omega^{2}}d\omega
\]
Eq.(\ref{u2floppy}) predicts a linear dependence of $\left\langle
u^{2}(T)\right\rangle $ upon $f.$ This result and the idea of an Einstein-like
mode for floppy modes are supported by the giant-softening of the $119Sn$
Lamb-Mossbauer factor in $(Ge_{0.99}Sn_{0.01})_{x}Se_{1-x}$ glasses
\cite{Boolchand2} as $x$ linearly decreases to $0$, \textit{i.e}., as the
glass gets more floppy. The mean square displacement at absolute zero or the
first moment of the vibrational density of states contains the information on
these floppy modes, and its variation $x$ nicely scales with the scattering
strength of the $5meV$ mode observed in inelastic scattering \cite{Boolchand2}.

Now we turn our attention to combine these results with the Lindemann
criteria. When $f=0$, we use eq.(\ref{gomega}) and the Lindemann criteria to
obtain $T_{g}$ at the rigidity threshold $f=0$,%
\begin{equation}
T_{g}(f=0)\approx\frac{0.1ma^{2}}{3k\left\langle \frac{1}{\omega^{2}%
}\right\rangle _{R}}\label{fzero}%
\end{equation}
For the variation of $T_{g}$ as a function of rigidity, we use the previous
expression to rewrite eq.(\ref{u2floppy}) in terms of $T_{g}(f=0)$, and then
we apply again the Lindemann criteria to obtain,
\begin{equation}
T_{g}(f)=\frac{T_{g}(f=0)}{1+\alpha f}\label{Tgf}%
\end{equation}
where $T_{g}(f)$ is the glass transition temperature when a fraction $f$ of
floppy modes is present. The parameter $\alpha$ is defined as,
\[
\alpha\equiv\frac{1}{\omega_{f}^{2}\left\langle \frac{1}{\omega^{2}%
}\right\rangle _{R}}-1\equiv\left(  \frac{\Theta_{R}}{\Theta_{f}}\right)
^{2}-1,
\]
and depends upon the ratio of two characteristic frequencies, since we define
the following frequency and temperature at the rigidity transition,
$\Theta_{R}/\hbar\equiv\omega_{R}\equiv\left\langle \frac{1}{\omega^{2}%
}\right\rangle _{R}^{-1/2}.$

Equation (\ref{Tgf}) predicts a decreasing glass temperature with the number
of floppy modes. Furthermore, by using the Maxwell mean field to express $f$
in terms of $<r>$, eq.(\ref{Tgf}) can be transformed into,%
\begin{equation}
T_{g}(\left\langle r\right\rangle )=\frac{T_{g}(\left\langle r\right\rangle
=2.0)}{1-\beta\left(  \left\langle r\right\rangle -2.0\right)  },\label{Gibbs}%
\end{equation}
where the constant $\beta$ is given by,%
\[
\beta=\frac{5\alpha}{2\alpha+6}.
\]
The value of $T_{g}(\left\langle r\right\rangle )$ obtained in eq.(\ref{Gibbs}%
) is exactly in the same form of the empirically modified Gibbs-DiMarzio
equation that has been used as a fitting law for many chalcogenide glasses,
where $\beta$ is a constant taken from the experimental data. Our approach
shows that this constant depends upon the ratio of two characteristic
frequencies or temperatures. It is worthwhile mentioning that the constant
experimental value of $\beta$ in the interval $2.0\leq\left\langle
r\right\rangle \leq2.4$ gives an extra support to the assumption made in
eq.(\ref{gomega}). For $\left\langle r\right\rangle >2.4$, the isocoordination
rule for $T_{g}$ is broken and $\beta$ is no longer constant for different
chemical compositions, as is also observed with the glass transition
temperature. Another advantage of the approach presented in this letter, is
that $\beta$ can be estimated from neutron scattering data or Lamb-Mossbauer
factor. Here we will use the neutron scattering data taken from reference
\cite{Kamitakahara}. We start by noting that at the rigidity transition, $f=0$
and from eq.(\ref{u2floppy}), $g_{I}(\omega)=g(\omega).$Thus, $\left\langle
\frac{1}{\omega^{2}}\right\rangle _{R}$ can be obtained from the experimental
data by calculating the second inverse moment of the normalized density of
states at $\left\langle r\right\rangle =2.4$. Following this procedure, we get
$\left\langle \frac{1}{\omega^{2}}\right\rangle _{R}\approx0.01986meV^{-2},$
with a characteristic frequency $\omega_{R}\approx7.0959meV$, and temperature
$\Theta_{R}\approx76%
%TCIMACRO{\U{b0}}%
%BeginExpansion
{{}^\circ}%
%EndExpansion
K.$Using that $\omega_{f}\approx5meV$, $\alpha$ has the approximate value
$1.014$, and finally $\beta\approx0.67$. The value from fitting the
experimental data of the glass transition has been obtained by many groups
\cite{Tatsumisago}\cite{Sreeram}\cite{GeorgievCR}, and produce the value
$\beta\approx0.72$ which is in very good agreement with our estimation from
neutron data. The calculation of $\beta$ can be improved by taking into
account that the floppy peak is not a delta function, but instead it has a
finite width, which will provide a small correction to the result. However,
even in this simple first approximation the results are encouraging.

In conclusion, we have explored the effects of floppy modes into the glass
transition temperature of glasses. Although these modes are not at zero
frequency and at high temperature they do not play a role in the specific
heat, they are fundamental to determine the glass transition temperature since
they enhance the square mean displacement. Then, by using the Lindemann
criteria and a very simple model of the vibrational density of states, we
recovered the empirically modified Gibss-DiMarzio equation that has been
widely used to account for the change in the transition temperature of
chalcogenide glasses. \ The constant that appear in this law depends upon two
parameters: the frequency of floppy modes and the second inverse moment of the
density of non-floppy modes. The results of this letter are in agreement with
Phillip%
%TCIMACRO{\U{b4}}%
%BeginExpansion
\'{}%
%EndExpansion
s original idea that glass forming tendency is enhanced at the rigidity
transition \cite{Phillips2}\cite{Phillips3}, as also thermodynamical
reversibility \cite{Boolchand1}, since it seems that there is less entropy due
to a smaller quadratic displacement.

\textbf{Acknowledgments.} This work was supported by DGAPA UNAM project
IN108502, and CONACyT -US National Science Foundation joint project 41538.

\end{document}